\documentclass[12pt]{iopart}
\usepackage{iopams}

\begin{document}

\title{On the Laplacian of $1/r$}
\author{D V Red\v zi\' c}

\address{Faculty of Physics, University of Belgrade, PO
Box 44, 11000 Beograd, Serbia} \eads{\mailto{redzic@ff.bg.ac.rs}}

\begin{abstract}
A novel definition of the Laplacian of $1/r$ is presented, suitable
for advanced undergraduates.
\end {abstract}

\section{Introduction}
Discussions of the Laplacian of $1/r$ generally start abruptly, {\it
in medias res}, by stating the relation
\begin {equation}
\nabla^2\frac {1}{r} = - 4\pi\delta^3(\bi r),
\end {equation}
where $r$ is the magnitude of radius vector $\bi r$ and
$\delta^3(\bi r)$ is the three-dimensional delta function, which is
then proved in various ways, clarifying thus its meaning. A glance
at equation (1) reveals, however, that the symbol $\nabla^2$
appearing in it can not have its ordinary, classical meaning of
$\bnabla \cdot \bnabla$, where, in Cartesian coordinates,
$$
\bnabla = \mbox {\pmb i}\frac {\partial}{\partial x} + \mbox {\pmb
j}\frac {\partial}{\partial y} + \mbox {\pmb k}\frac
{\partial}{\partial z}
$$
is Hamilton's operator nabla (cf, e.g., Red\v zi\' c \cite{DVR}),
since the classical expression $\nabla^2(1/r)$ vanishes for $r \neq
0$ and is not defined at $r = 0$. Therefore, instead of the familiar
form (1), in this note we will use a less confusing notation
\cite{Estrada,VH}
\begin {equation}
\bar{\nabla}^2\frac {1}{r} = - 4\pi\delta^3(\bi r);
\end {equation}
the expression $\bar{\nabla}^2(1/r)$ we will call the generalized
(distributional) Laplacian of $1/r$ and try to fathom its meaning.
Let us briefly review some typical proofs of (2).

The well-known way to demonstrate (2) is to regularize $1/r$ in
terms of a parameter $a$ so that the regularized function is
well-behaved everywhere for $a \neq 0$. Then verification of (2)
consists in showing that in the limit $a \rightarrow 0$, $-1/4\pi$
times the Laplacian of the regularized function is a representation
of the three-dimensional delta function $\delta^3(\bi r)$. For
example, regularizing $1/r$ as $1/\sqrt {r^2 + a^2}$, Jackson
\cite{JDJ} shows that
\begin {equation}
\bar{\nabla}^2\frac {1}{r} = \lim_{a \rightarrow 0}\nabla^2\frac
{1}{\sqrt {r^2 + a^2}} = - 4\pi\delta^3(\bi r);
\end {equation}
the limit here is the weak limit (cf, e.g., \cite{VSV}). A more
sophisticated method of proving (2) would be to derive first the
generalized second-order partial derivatives of $1/r$ with respect
to Cartesian coordinates \cite{VH,CPF}.\footnote [1]{A typical {\it
informal} derivation of (2) (cf, e.g., \cite {SMB}) starts from
relation
$$
\int\nabla^2\frac {1}{r}\rmd^3r = \int\bnabla \cdot\left (-\frac
{\bi r}{r^3}\right )\rmd^3r,
$$
which, using the divergence theorem, equals $-4\pi =
-4\pi\int\delta^3(\bi r)\rmd^3r$. However, the use of the divergence
theorem is not legitimate here, since the function $-\bi r/r^3$ is
singular at $r = 0$, and the result
$$
\int\nabla^2\frac {1}{r}\rmd^3r = -4\pi,
$$
is incorrect since the left-hand side of the last equation equals
zero.}

A generalization of (2)
\begin {equation}
\bar{\nabla}^2\frac {1}{|\bi r - \bi r'|} =  - 4\pi\delta^3(\bi r -
\bi r'),
\end {equation}
obtained simply by a different choice of the origin, can be
demonstrated either via the regularization procedure\footnote [2]
{Jackson \cite{JDJ}, in fact, shows the validity of a generalization
of (3)
$$
\bar{\nabla}^2\frac {1}{|\bi r - \bi r'|} = \lim_{a \rightarrow
0}\nabla^2\frac {1}{\sqrt {(\bi r - \bi r')^2 + a^2}} = -
4\pi\delta^3(\bi r - \bi r').
$$}
or by employing a well-known electrostatic argument. Namely, it can
be shown that the `potential'
\begin {equation}
\int \varrho (\bi r')\frac {1}{|\bi r - \bi r'|}\rmd^3r',
\end {equation}
where the `density' $\varrho (\bi r')$ plays the role of a
well-behaved `test' function, satisfies Poisson's equation
\begin {eqnarray}
\nabla^2\int \varrho (\bi r')\frac {1}{|\bi r - \bi r'|}\rmd^3r' &=
\bnabla \cdot\int \varrho (\bi r')\bnabla\frac {1}{|\bi r - \bi r'|}\rmd^3r'  \nonumber \\
&= -4\pi \varrho (\bi r),
\end {eqnarray}
by making use of the divergence theorem and Gauss's theorem from
electrostatics \cite{MF,VH2}. Since the right-hand side of equation
(6) can be written as
\begin {equation}
-4\pi\int \varrho (\bi r')\delta^3(\bi r - \bi r')\rmd^3r',
\end {equation}
it follows that the operator $\nabla^2$ can enter an integral of the
form (5) under proviso that `during entrance' it converts into the
generalized operator $\bar{\nabla}^2$ whose action on the function
$1/|\bi r - \bi r'|$ is defined by equation (4).

On the other hand, the representation formula for a well-behaved
scalar function of position $\Phi$
\begin {equation}
\Phi (\pmb r) = -\frac {1}{4\pi }\int_{V} \frac {\nabla'^2\Phi (\pmb
r')}{\cal R}\rmd^3r' + \frac {1}{4\pi}\oint_{S} \left[\frac {1}{\cal
R}\frac {\partial \Phi}{\partial n'} - \Phi \frac {\partial
}{\partial n'}\left (\frac {1}{\cal R}\right ) \right ]\rmd S' \, ,
\end {equation}
where the point $\bi r$ is {\it within} the volume $V$ and $\cal R
\equiv |\bi r - \bi r'|$, is obtained from Green's second identity
\begin {equation}
\fl\int_{V - V_\varepsilon}\left (\frac {1}{\cal R}\nabla'^2\Phi
(\pmb r') - \Phi (\pmb r')\nabla'^2\frac {1}{\cal R}\right )\rmd^3r'
= \left (\oint_{S} + \oint_{S_\varepsilon}\right ) \left[\frac
{1}{\cal R}\frac {\partial \Phi}{\partial n'} - \Phi \frac {\partial
}{\partial n'}\left (\frac {1}{\cal R}\right ) \right ]\rmd S' \, ,
\end {equation}
where the volume $V - V_\varepsilon$ is obtained by excluding a
small ball of radius $\varepsilon$ and centre at $\bi r' = \bi r$
from the volume $V$, and $S_\varepsilon$ is the surface of the ball,
taking the limit $\varepsilon \rightarrow 0$ (cf, e.g. \cite {JAS}).
In this way, the singularity of the function $1/\cal R$ at $\bi r' =
\bi r$ is managed with, making possible the use of Green's second
identity. As is well known, this classical procedure provides
grounds for replacing the term $\nabla'^2(1/\cal R)$ in (9) by its
generalized counterpart $\bar{\nabla}'^2(1/\cal R)$ defined by (4),
extending of course at the same time the restricted
(singularity-free) volume of integration $V - V_\varepsilon$ to the
whole region $V$ and removing the integral over the surface
$S_\varepsilon$.\footnote [3] {To put it rigorously, in this case
relation
$$
\lim_{\varepsilon \rightarrow 0}\left \{\int_{V - V_\varepsilon}
\Phi (\pmb r')\nabla'^2\frac {1}{\cal R}\rmd^3r' +
\oint_{S_\varepsilon} \left[\frac {1}{\cal R}\frac {\partial
\Phi}{\partial n'} - \Phi \frac {\partial }{\partial n'}\left (\frac
{1}{\cal R}\right ) \right ]\rmd S'\right \} = -4\pi\Phi (\pmb r)
$$
serves as a definition of $\bar {\nabla}^2(1/\cal R)$ (cf, e.g.,
\cite {VSV}), since
$$
-4\pi\Phi (\pmb r) = -4\pi\int\Phi (\pmb r')\delta^3 (\bi r' - \bi
r)\rmd^3r'.
$$}
Thus, employing of the generalized Laplacian of the function $1/|\bi
r - \bi r'|$ instead of the classical one in regions that contain
the singular point $\bi r' = \bi r$, replaces lengthy procedures
based on classical analysis, providing a shortcut to the correct
final results.

In this note, an alternative definition of the generalized Laplacian
of $1/r$ will be presented. Taking into account the ubiquity and
importance of this somewhat tricky concept, the alternative
derivation of relation (2) could perhaps be of some pedagogical
interest.

\section{Integral versus differential definitions of classical operators}
A perusal of the literature reveals that a common feature of
discussions of the Laplacian of $1/r$ is that, right from the
outset, the Laplacian is identified with $\nabla^2 \equiv \bnabla
\cdot \bnabla$. That is, in the {\it primary} definition of the
Laplacian of a scalar field, which is the divergence of the gradient
of the field, both the divergence and gradient operators are
understood according to their differential definitions, $\bnabla
\cdot$ and $\bnabla$, respectively, in accord with common practice
of defining familiar operators in vector analysis by differential
operations (cf, e.g., \cite{DVR,DJG}). However, as is well known,
there exists also an alternative way of defining the classical
operators by means of integral operations (cf, e.g.,
\cite{MF,AS,KORN}). While the integral and differential definitions
are equivalent in the case of differentiable fields, the former
definition provides physical insight and computational convenience,
and thus should be preferred, as Sommerfeld suggested \cite {AS}.
Moreover, it appears that the integral definitions of classical
operators are more to the point than the differential ones when
applied to a singular point of the field. Let us examine the last
point in some detail.

As a simple illustration, consider the Laplacian of $1/r$ at the
singular point $r = 0$. Using the differential definitions,
$\nabla^2(1/r)$ is not defined at $r = 0$. On the other hand, using
the integral definition of the divergence of a vector field $\bi A$
at a point P,
\begin {equation}
\mbox {div}\bi A = \lim_{\tau \rightarrow 0}\frac {1}{\tau}\oint \bi
A \cdot \rmd\bi S,
\end {equation}
where the flux of $\bi A$ is through a closed surface surrounding P
and $\tau$ is the enclosed volume, and setting $\bi A = \mbox
{grad}(1/r) = - \bi r/r^3$, we obtain that
\begin {equation}
\mbox {div\,grad}(1/r) = -4\pi\infty = -\infty,
\end {equation}
at $r = 0$. Since infinite value of a function at a point is not
permitted in classical analysis, the two definitions seem to be in a
dead heat. However, this is not so, as the following argument will
show.

Calculate the average value of the Laplacian of $1/r$ over the
volume of a ball of radius $\varepsilon$ and centre at $r = 0$.
Using the differential definition of the Laplacian we obtain
\begin {equation}
\langle\nabla^2(1/r)\rangle = \frac
{1}{(4/3)\pi\varepsilon^3}\int_{V_{\varepsilon}}
\nabla^2(1/r)\rmd^3r = 0,
\end {equation}
where $\langle...\rangle$ stands for the average value, and
$V_{\varepsilon}$ is the volume of the ball. Once again, the same
result is obtained using the integral definition of the Laplacian,
\begin {equation}
\langle\mbox{div\,grad}(1/r)\rangle = \frac
{1}{(4/3)\pi\varepsilon^3}\int_{V_{\varepsilon}} \mbox{div}(-\bi
r/r^3)\rmd^3r = 0,
\end {equation}
since according to both definitions the Laplacian of $1/r$ vanishes
for $r\neq 0$, and the value of integral is not affected by a
countable number of singular points of the integrand.

Inspection of equation (10) reveals, however, that a reasonable
physical definition of the {\it volume average} of the divergence of
a vector field $\bi A$ over a volume $\tau$ would be
\begin {equation}
\langle\mbox {div}\bi A\rangle = \frac {1}{\tau}\oint_{S_{\tau}} \bi
A \cdot \rmd\bi S,
\end {equation}
where $S_{\tau}$ is the surface of the volume $\tau$, rather than
the standard definition
\begin {equation}
\langle\mbox {div}\bi A\rangle = \frac {1}{\tau}\int_{\tau}
\mbox{div}\bi A \rmd^3r.
\end {equation}
Namely, while both definitions yield the same result when the field
$\bi A$ is differentiable over the volume $\tau$, they give
different results when $\bi A$ is singular at a point inside $\tau$.
For example, if $\bi E$ is the electrostatic field of a point charge
$q$ located at the origin, and $\tau$ is a volume containing the
origin, $\langle\mbox {div}\bi E\rangle$ would be equal to
$q/\tau\epsilon_0$ according to definition (14). On the other hand,
according to definition (15), $\langle\mbox {div}\bi E\rangle$ would
be equal to zero, since $\mbox {div}\bi E$ is zero for $r \neq 0$
and, classically, is not defined at $r = 0$. Keeping in mind that
the divergence of a vector field is a measure of the strength of a
source or sink of field lines, it is clear that the standard
definition (15) would yield the absurd result that the average
strength of a point source over a volume containing the source is
zero.\footnote [4] {Note that for a given field, a source of field
lines and a source of the field need not necessarily coincide. For
example, the electric field lines of a point charge which is forever
in uniform motion emanate from the present position of the charge,
whereas the source of the field is the charge at the retarded
positions (cf, e.g.\cite{DJG,APF}).} Thus, the integral definition
(14) of the average divergence not only implies the integral
definition (10) of the divergence at a point, but also is adequate
for regions that contain singularities (point sources or sinks) of
field lines.\footnote [5] {Recall that a valid definition, {\it
inter alia}, should be adequate (`{\it definitio sit adequata}').
Recall also that there are other cases when the average value of a
physical quantity can not be defined via an integral of the {\it
classical} local values of the quantity. For example, the average
velocity of a particle, is primarily given by $\langle\bi v\rangle =
\triangle\bi r/\triangle t$, where $\triangle\bi r$ is a
displacement of the particle during a time interval $\triangle t$,
and not by $\langle\bi v\rangle = (1/\triangle t)\int_0^{\triangle
t}\bi v(t)\rmd t$. (The two definitions are equivalent only if $\bi
r$ is a differentiable function of $t$.) Similarly, the average
charge density over a volume $\triangle V$ is primarily
$\langle\varrho\rangle = \triangle Q/\triangle V$, where $\triangle
Q$ is a charge inside $\triangle V$, and not $\langle\varrho\rangle
= (1/\triangle V)\int_{\triangle V}\varrho(\bi r)\rmd^3r$, if
$\varrho(\bi r)$ is the {\it classical} local charge density. Of
course, if $\bi v(t)$ and $\varrho(\bi r)$ are described by
generalized functions (distributions), such as the Dirac delta
function (which is the case when $\bi v(t)$ changes abruptly during
$\triangle t$, or there are point charges inside the volume
$\triangle V$), no discrepancy arises between the two kinds of
definitions.}

\section{A novel definition of the Laplacian of $1/r$}

The above analysis provides the opportunity of introducing an
alternative definition of the generalized Laplacian of $1/r$.

As a first step, define the $\varepsilon$-Laplacian of $1/r$, $\mbox
L_{\varepsilon}(1/r)$, in terms of a parameter $\varepsilon$ as
\begin{equation}
\mbox L_{\varepsilon}\frac {1}{r} \equiv
\mbox{div}_{\varepsilon}\mbox{grad}\frac {1}{r}\equiv \left\{
  \begin{array}{cll}
[\oint_{r = \varepsilon}(-\bi r/r^3)\cdot \rmd\bi
S]/(4/3)\pi\varepsilon^3,& {\rm if} & r < \varepsilon\\
\mbox {div\,grad}(1/r) = 0, & {\rm if} & r > \varepsilon.
  \end{array}
  \right.
\end{equation}
Thus, for $r < \varepsilon$, $\mbox L_{\varepsilon}(1/r)$ is the
average divergence (as defined by the integral definition (14)) of
the gradient of $1/r$ over the volume of a ball of radius
$\varepsilon$ and centre at $r = 0$; for $r
> \varepsilon$, $\mbox L_{\varepsilon}(1/r)$ is the classical Laplacian of $1/r$. Definition (16) implies that
\begin{equation}
\mbox L_{\varepsilon}\frac {1}{r} = -\frac
{3}{\varepsilon^3}\Theta(\varepsilon - r)
\end{equation}
where $\Theta(x)$ is the Heaviside step function,
\begin{equation}
 \Theta(x) =\left\{
  \begin{array}{lll}
  0,& {\rm if} & x \leq 0 \\
  1, & {\rm if} & x > 0.
  \end{array}
  \right.
\end{equation}
When $\varepsilon \rightarrow 0$, from equation (17) we obtain
\begin{equation}
 \lim_{\varepsilon \rightarrow 0}\mbox L_{\varepsilon}\frac {1}{r}  =\left\{
  \begin{array}{cll}
  -\infty,& {\rm if} & r = 0 \\
  0, & {\rm if} & r \neq 0.
  \end{array}
  \right.
\end{equation}
On the other hand, integrating the product of $\mbox
L_{\varepsilon}(1/r)$ and a well-behaved `test' function $f(\bi r)$,
\begin{equation}
\int\left[\mbox L_{\varepsilon}\frac {1}{r}\right]f(\bi r)\rmd^3r =
-\int_{r \leq \varepsilon}\frac {3}{\varepsilon^3}f(\bi r)\rmd^3r,
\end{equation}
using expansion of $f(\bi r)$ in a Taylor series around $\bi r = 0$,
and taking the limit $\varepsilon \rightarrow 0$ yields

\begin {eqnarray}
\lim_{\varepsilon \rightarrow 0}\int\left[\mbox L_{\varepsilon}\frac
{1}{r}\right]f(\bi r)\rmd^3r  & = -\lim_{\varepsilon \rightarrow
0}\int_0^{\varepsilon}\frac {3}{\varepsilon^3}
\left [f(0) + \frac {r^2}{6}\nabla^2 f + ... \right ]4\pi r^2\rmd r \nonumber \\
&= -4\pi f(0).
\end {eqnarray}
As is well known, result (21) can be expressed as

\begin {equation}
\bar {\mbox L}\frac {1}{r} = {\rm w}\!\!\lim_{\varepsilon
\rightarrow 0}\mbox L_{\varepsilon}\frac {1}{r} = - 4\pi\delta^3(\bi
r);
\end {equation}
where a more suitable notation $\bar {\mbox L}$ for the generalized
Laplacian is now used instead of $\bar {\nabla}^2$ and wlim stands
for the weak limit (cf, e.g., \cite{VH,VSV}).\footnote [6] {As was
pointed out in Section 2, the standard practice of using
automatically $\nabla^2$ for the classical Laplacian may be
misleading.} Equation (22) is tantamount to equation (3), providing
another definition of the generalized Laplacian of $1/r$.\footnote
[7] {Recall that the 2D analogue of equation (4) reads
$$
\bar{\nabla}^2\ln|\bi s - \bi s'| =  2\pi\delta^2(\bi s - \bi s'),
$$
where $\bi s$ and $\bi s'$ are 2D radius vectors and $\delta^2(\bi s
- \bi s')$ is the 2D delta function \cite{JDJ,MF,JCR}, which setting
$\bi s' = 0$ yields
$$
\bar{\nabla}^2\ln s =  2\pi\delta^2(\bi s),
$$

\noindent which is the 2D analogue of equation (3). To  prove the
last relation, instead of regularizing $\ln s$ in terms of a
parameter $a$ as $\ln\sqrt{s^2 + a^2}$, the alternative 2D
definition of the generalized Laplacian of $\ln s$,
$$
\bar{\mbox L}\ln s =  2\pi\delta^2(\bi s),
$$
can be introduced, following a 2D procedure analogous to the 3D
procedure described above.}

Discuss now briefly a closely related problem of defining
analogously the generalized charge density for a {\it point charge}
$q$ located at the origin. Obviously, the corresponding
$\varepsilon$-charge density should be defined as
\begin{equation}
\varrho_{\varepsilon}(\bi r) = \left\{
  \begin{array}{cll}
q/(4/3)\pi\varepsilon^3,& {\rm if} & r < \varepsilon\\
 \varrho (\bi r) = 0, & {\rm if} & r > \varepsilon,
  \end{array}
  \right.
\end{equation}
which can be recast into
\begin{equation}
\varrho_{\varepsilon}(\bi r) = \frac
{3q}{4\pi\varepsilon^3}\Theta(\varepsilon - r).
\end{equation}
Passing details, we give the final result
\begin {equation}
\bar{\varrho}(\bi r) = {\rm w}\!\!\lim_{\varepsilon \rightarrow
0}\varrho_{\varepsilon}(\bi r) = q\delta^3(\bi r).
\end {equation}
where $\bar{\varrho}(\bi r)$ is the generalized charge density. Note
that $\lim_{\varepsilon \rightarrow 0}\int\varrho_{\varepsilon}(\bi
r)\rmd^3r = q$, whereas $\int [\lim_{\varepsilon \rightarrow
0}\varrho_{\varepsilon}(\bi r)]\rmd^3r = 0$. Thus, the volume charge
density of a point charge $q$ located at $r = 0$ is naturally
described by $\varrho_{\varepsilon}(\bi r)$ and its weak limit
$\bar{\varrho}(\bi r)$, in perfect analogy with the volume flux
density of the flux of grad$(1/r)$, which is naturally described by
$\mbox L_{\varepsilon}(1/r)$ and its weak limit $\bar {\mbox
L}(1/r)$.

In the same way we can define the generalized divergence of the
electric field $\bi E(\bi r,t)$  of a point charge $q$ that is
moving on a specific trajectory $\bi r_q(t)$. Starting from the
corresponding $\varepsilon$-divergence
\begin{equation}
\mbox {div}_{\varepsilon}\bi E(\bi r,t) \equiv
 \left\{
  \begin{array}{cll}
[\oint_{\xi = \varepsilon}\bi E\cdot \rmd\bi
S]/(4/3)\pi\varepsilon^3,& {\rm if} & \xi < \varepsilon\\
\mbox {div}\bi E = 0, & {\rm if} & \xi > \varepsilon,
  \end{array}
  \right.
\end{equation}
where $\xi = |\bi r - \bi r_q(t)|$, using Gauss's law in its
integral form, $\oint_{\xi = \varepsilon}\bi E\cdot \rmd\bi S =
q/\epsilon_0$, we obtain that $\bi E(\bi r,t)$ must satisfy equation
\begin{equation}
\overline {\mbox {div}}\bi E (\bi r,t) = \frac {q\delta(\bi r - \bi
r_q(t))}{\epsilon_0} = \frac {\bar{\varrho}(\bi r, t)}{\epsilon_0}
\end{equation}
where
\begin{equation}
\overline {\mbox {div}}\bi E (\bi r,t) = {\rm
w}\!\!\lim_{\varepsilon \rightarrow 0}\mbox {div}_{\varepsilon}\bi
E(\bi r,t)
\end{equation}
is the generalized divergence of $\bi E (\bi r,t)$. This is
consistent with a general {\it sine qua non} that when the sources
of a field are idealized as point, line or surface distributions of
charge and/or current, described by generalized functions,
generalized operators must be employed in the field or potential
equations instead of the classical ones.\footnote [8] {An
instructive illustration of the possible pitfalls of using the delta
function in the context of solving a classical 2D electrodynamic
problem is presented in \cite{DRVH}.}

\section{Conclusions}

We presented a novel definition of the generalized Laplacian of
$1/r$, avoiding regularization of $1/r$ \cite{VH,JDJ},
`electrostatic' procedure \cite{MF,VH2}, or Green's second identity
\cite{VSV}. The definition is constructed employing the integral
definition of the divergence instead of the differential one; thus,
the usual notation $\bar {\nabla}^2$ is replaced by a less
suggestive one $\bar{\mbox L}$. It is shown that the Laplacian of
$1/r$ can be naturally construed as the volume flux density of the
flux of grad$(1/r)$, in the same way as the volume charge density of
a point charge located at the origin, introducing a reasonable
generalized density. We believe that our analysis provides a simple
and insightful alternative to the earlier discussions of the
concept, clarifying thus its meaning.

\section*{Acknowledgments}
I thank Vladimir Hnizdo for illuminating correspondence and Du\v san
Georgijevi\' c for helpful comments on an earlier draft.

\end{document}